# Å-Indentation for non-destructive elastic moduli measurements of supported ultra-hard ultra-thin films and nanostructures

Filippo Cellini[1,2], Yang Gao[2,4], Elisa Riedo*[1,2,3,4]

[1] Tandon School of Engineering, New York University, Brooklyn, NY 11201, USA

[2] Advanced Science Research Center, CUNY Graduate Center, New York, NY 10031, USA

[3] Physics Department, City College of New York, CUNY, New York, NY 10031, USA

[4] School of Physics, Georgia Institute of Technology, Atlanta, GA 30332, USA

*Corresponding Author, e-mail: elisa.riedo@nyu.edu

*Abstract* — **During conventional nanoindentation measurements, the indentation depths are usually larger than 1-10 nm, which hinders the ability to study ultra-thin films (< 10 nm) and supported atomically thin two-dimensional (2D) materials. Here, we discuss the development of modulated Å-indentation to achieve sub-Å indentation depths during force-indentation measurements while also imaging materials with nanoscale resolution. Modulated nanoindentation (MoNI) was originally invented to measure the radial elasticity of multi-walled nanotubes. Now, by using extremely small amplitude oscillations (<< 1 Å) at high frequency, and stiff cantilevers, we show how modulated nano/Å-indentation (MoNI/ÅI) enables non-destructive measurements of the contact stiffness and indentation modulus of ultra-thin ultra-stiff films, including CVD diamond films (modulus ~1000 GPa), as well as the transverse modulus of 2D materials. Our analysis demonstrates that in presence of a standard laboratory noise floor, the signal to noise ratio of MoNI/ÅI implemented with a commercial atomic force microscope (AFM) is such that a dynamic range of 80 dB — achievable with commercial Lock-in amplifiers — is sufficient to observe superior indentation curves, having indentation depths as small as 0.3 Å, resolution in indentation < 0.05 Å, and in normal load < 0.5 nN. Being implemented on a standard AFM, this method has the potential for a broad applicability.**

# I. INTRODUCTION

Nanoindentation has been continuously applied in the last two decades to investigate the mechanical properties of materials at the nanoscale. The main advantages of this technique are the extremely small imposed deformation, of the order of few hundred nanometers, the small loading force, ranging from few mN to μN, and the small sample size required for testing, with sample surfaces of less than a few microns squared [1, 2]. Nanoindentation has enabled for the first time the direct measurement of the stiffness and hardness of thin films and coatings, with a better resolution compared to previous testing methodologies. Nanonindentation has been primarily adopted in the scientific and industrial practice in the study of metal alloy thin films [3] and inorganic crystalline thin films [4], but several studies have also explored its application to biomaterials, such as bones [5] and DNA [6], polymeric films [7], and colloidal crystals [8].

The nanoindentation technique most widely adopted by the scientific community is the one described in references [1, 2, 9]. This technique has also been usually referred to as continuous stiffness measurement (CSM) [1]. In its fundamental incarnation, this methodology employs a nanoindenter of known geometry, usually a tetragonal tip or Berkovich indenter, and size, which is pressed against the sample using an actuation system controlled by a magnetic coil [1, 2, 9]. Characteristic indentation depths usually range from a few hundred nanometers to several microns. A small residual deformation is usually observed after indentation, whereby a micrometer size indent is left on the surface of the specimen after the test. During the CSM experiment, the mechanical stiffness of the material is continuously measured by superimposing to the main indentation force, a small periodic oscillation at a known frequency below 100 Hz. The periodic force and indenter displacements are acquired using a Lock-in amplifier, and employed to compute the contact stiffness [1, 2, 9].

Several studies have been recently devoted to develop the classical nanoindentation technique in order to improve its resolution at small scales, reduce the uncertainty associated to the measurement chain, and remove the necessity to visually inspect the residual indent after testing. For example, a high precision nanoindentation instrument based on an instrumented indentation testing (IIT) layout has been developed in reference [10]. This instrument is able to conduct nanoindentation measurements with noise floor of 2 μN, displacement resolution of 0.4 nm, and uncertainty in the indentation depth of less than 10 nm. An ultra nanoindentation tester has been presented in reference [11], whereby indentation measurements of few nanometers depth are performed using a Berkovich indenter with a remarkable resolution of 1 nN in force and 0.3 pm in displacement. An important contribution to the field is the recent development of second harmonic detection methods, which are discussed in references [12, 13]. In this class of methods, the normal displacement is not used to identify the mechanical properties, thereby reducing the need of accurate detection of the contact between the indenter and the sample.

While nanoindetation techniques have been successfully employed to characterize a wide range of materials, there are some fundamental limits that hinder their application when the size of the structures under investigation goes below a certain scale (a comprehensive description of the measurement process is found in [14]). More specifically, indentation depths usually employed during conventional nanoindentation are of the order of at least 20-100 nanometers, which is the same order of magnitude or larger than the thickness of ultra-thin films (< 10 nm) increasingly studied for semiconductor and energy applications [15, 16], or the thickness of nanowires [17] and atomically thin two-dimensional (2D) materials (~1-5 nm) [18-20]. Indentation depths that are comparable, if not larger, than the thickness of the film may compromise the accuracy of the measurement, substantially increasing the contribution of the substrate to the mechanical response and eventually disrupting and damaging the film. In addition, classical nanonindentation usually produces a permanent deformation of the material (the residual indent), which might result in an irreversible modification of properties of the thin film beyond its mechanical stiffness, such as its electrical or thermal conductivity at the nanoscale [21, 22].

Nanomechanical measurements based on the Atomic Force Microscopy (AFM) have been explored as an alternative to classical nanoindentation [23-25]. While some scholars still question the ability of AFM-based measurements to replace classical nanoindentation in the mechanical characterization of surfaces, several studies have demonstrated the robustness of these methodologies for measuring the mechanical properties of materials [24, 26]. By leveraging the high speed scanning/imaging capabilities of the AFM, research efforts have been traditionally devoted to develop imaging techniques for qualitative estimation of mechanical properties [27, 28], while quantitative estimation has been achieved employing an external calibration reference [29-31]. For example, a bimodal AM-FM imaging method is employed in reference [31] to measure the stiffness of materials ranging from polymers (few hundred MPa) to titanium and silicon (~100 GPa). In addition, new methodologies are continuously under study [32, 33]. For instance, a multifrequency force spectroscopy method has been recently presented in reference [33], whereby mechanical modulus and viscoelastic properties of soft materials are successfully identified with great accuracy.

Herein, we discuss the experimental details and application of a novel sub-Å resolution indentation method based on AFM and Modulated Nano-indentation (MoNI), which enables elasticity measurements of ultra-thin and ultra-hard materials (100-1000 GPa) with very shallow indentation depths — as small as 0.3 Å— and with force and displacement/indentation resolution of < 0.5 nN and < 0.05 Å, respectively. We call this method "Å-*indentation*" (ÅI) to better reflect its extremely high resolution (sub-Å) and outstanding ability to probe sample surface phenomena at Å-scale depth. The technique is based on an approach similar to well-established CSM methodologies, whereby a high frequency oscillating force is superimposed to the main force applied during the indentation cycle using a commercial AFM system. Differently from traditional CSM, MoNI/ÅI employs extremely small amplitude oscillations (<< 1 Å), higher frequency, and force detection systems based on laser detection AFM methods. This technique, however simple, is shown to provide accurate estimations. More specifically, our analysis demonstrates that in presence of a standard laboratory noise floor, the signal to noise ratio of MoNI/ÅI implemented with a commercial atomic force microscope (AFM) is such that a dynamic range of 80 dB —achievable with commercial Lock-in amplifiers— is sufficient to observe superior indentation curves, with indentation depths as small as 0.3 Å, resolution in indentation < 0.05 Å and in normal load < 0.5 nN. The traditional definition of dynamic reserve is the ratio of the largest tolerable noise signal to the full scale signal, expressed in dB. In our analysis the full scale indentation signal for ultra-stiff films (~1000 GPa stiffness) is as low as 2 pm, then a dynamic reserve of 80 dB means cantilever vibration noise as large as 20 nm (80 dB greater than full scale) can be tolerated at the input without overload. In what follows, we show how the detection of the indentation signal can be performed above the noise level using a commercial Lock-in amplifier, whose dynamic range is 100 dB [34].

MoNI/ÅI has recently found applications in the characterization of the mechanical properties of the transverse mechanical stiffness of supported 2D materials, and in particular supported epitaxial graphene [19, 35]. The enhanced resolution of MoNI/ÅI has enabled for the first time the direct measurement of the inter-layer stiffness of few layer thick graphene and graphene oxide supported films [19] and has led to the discovery of the room-temperature diamondization of epitaxial bi-layer graphene on silicon carbide [35, 36]. These measurements have been possible only due to the extremely high spatial resolution of MoNI/ÅI. Further, the possibility of informing the MoNI/ÅI measurements with topographic AFM imaging allows resolution and testing of small features on the sample. For example, MoNI/ÅI has been employed to detect and probe nanofilaments of Q-carbon of 30 nm depth and 200 nm width [37]. This level of lateral resolution is of course extremely hard to achieve with traditional nanoindentation techniques.

In this paper, we discuss in details the instrumentation, calibration, and procedure employed for MoNI/ÅI, as well as the analysis performed to measure the contact stiffness and reconstruct the indentation curves. More specifically, in Section II we present the experimental apparatus and the procedure for MoNI/ÅI. We also present the calculations employed to compute the stiffness of the contact, the indentation curves, and the elastic modulus of the material. Further, we describe the calibration procedure, the estimation and prediction of the noise levels, we provide an example of the input and output data in a typical MoNI/ÅI experiment, and we estimate the uncertainty associate to the identification of the indentation modulus. In Section III, we report on the results obtained in MoNI/ÅI experiments conducted on reference materials, namely CVD diamond, sapphire, zinc oxide, and silicon oxide as well as the results obtained in atomically thin graphene and graphene oxide films on silicon carbide. We also discuss the unique features of MoNI/ÅI that have allowed these unprecedented measurement of the inter-layer/transverse elasticity of 2D films and the phase transformation of bilayer epitaxial graphene on silicon carbide into diamene. The article concludes in Section IV with a summary of the main results.

## II. MATERIALS AND METHODS

### A. Experimental Apparatus and Procedure

The setup for MoNI/ÅI experiments is described in the schematic displayed in Figure 1. The experimental apparatus is composed of an Agilent PicoPlus AFM, which is interfaced with a Stanford Research Systems SR830 DSP Lock-in amplifier, and a voltage divider assembled in-house. In the schematic, the oscillating voltage output of the Lock-in amplifier is fed through the divider to the piezotube controlling the displacement along the z-axis of the AFM cantilever. The raw deflection measured by the AFM photodetector is transmitted back through the AFM controller to the phase-sensitive detector of the Lock-in amplifier. Raw data from the Lock-in recorded either by using the AFM Software or a National Instruments Data Acquisition board are employed to compute the mechanical properties of the material. A detailed description of the methodology employed to compute the mechanical properties of the sample surface from raw data is reported in Section II B. The AFM cantilevers typically employed in MoNI/ÅI experiments have nominal spring constant in the range between 40 and 230 N/m, with the first resonant frequency between 270 and 600 kHz. Two different types of AFM probes have been employed in MoNI/ÅI experiments with comparable results, traditional Silicon tips and diamond-coated silicon tips, with diamond-coated tips being the preferred choice in order to minimize the effect of tip wear over repeated measurements. Accurate calibration of the tip and cantilever is required before performing MoNI/ÅI. A detailed description of the calibration procedure is reported in Section II C.

During the MoNI/ÅI experiments, as mentioned above, a Lock-in amplifier is used to generate a sinusoidal voltage signal at a fixed frequency, $\Delta V^{\text{Lock-in}}(t) = |\Delta V^{\text{Lock-in}}|\sin(2\pi f\, t + \varphi)$, where the voltage $\Delta V^{\text{Lock-in}}$ is usually in the range 4-8 mV, the oscillation frequency $f \approx 991$ Hz, and $\varphi$ is a phase shift. The voltage generated by the Lock-in amplifier is then reduced using a divider, with a diving factor $D$ ranging between 1 and 1000. The voltage $\Delta V^{\text{Lock-in}}(t)/D$ is applied to the piezotube of the AFM cantilever holder to control the cantilever/tip displacement along the z-axis. A tension of a few tenths of mV applied by the Lock-in amplifier results in an oscillation of the piezotube and the rigidly connected AFM cantilever holder of few Angstroms. For example, by considering a tension of $\Delta V^{\text{Lock-in}} = 4$ mV RMS, a dividing factor D = 10, and a piezo coefficient calibration of C= 1.8 Å/mV RMS (see Figure 5(a) in Section II), we get an oscillatory amplitude of the piezotube of $\Delta z_{\text{piezo}} = 0.7$ Å. This extremely small displacement can be measured by means of the four-quadrant AFM photodetector, whose output deflection signal is then read using the phase-sensitive detector of the Lock-in amplifier, which allows reconstruction of signals with intensities way below the noise floor of the AFM. In Section II D, we discuss the effect of noise in MoNI/ÅI and the phase-sensitive detection operated by the Lock-in amplifier.

In a typical MoNI/ÅI experiment, the tip is initially positioned in contact with a surface, and a certain initial load voltage is applied to the piezotube to further generate a contact between the AFM tip and the surface with a constant force, $F_0^z$, between the tip and the sample. When the MoNI/ÅI experiment starts, the total displacement of the piezotube is composed of two components: a constant displacement due to the load voltage applied to the piezotube and imposed by the AFM controller ($V_0^{\text{AFM}}$ in Figure 1), and a superimposed sinusoidal oscillating deflection of very small amplitude due to the sinusoidal voltage signal ($\Delta V^{\text{Lock-in}}(t)/D$ in Figure 1) generated by the Lock-in amplifier at fixed frequency and also applied to the piezotube (corresponding to a vertical oscillating displacement of the piezotube $\Delta z_{\text{piezo}}$). After the tip is brought into contact with the surface, the load voltage is slowly decreased to progressively reduce the contact force between the tip and the sample. This change is quasi-static and driven at a rate (<0.002 V/s) much slower than the oscillation frequency of the sinusoidal signal (991 Hz, approx. 0.5 V/s) generated by the Lock-in amplifier. During the unloading phase, the feedback loop of the AFM controls the load force $F_0^z$ while the small oscillations, too small to be read by the AFM controller, are applied to the piezotube. This ensures that the load $F_0^z$ is maintained and corrected by possible thermal drifts. Therefore, for each fixed $F_0^z$, the Lock-in records the cantilever deflection signal $F_{\text{tot}} = F_0^z + \Delta F(F_0^z) \sin(2\pi f\, t + \varphi)$ to compute $\Delta F(F_0^z)$ and thus obtain the local elasticity of the material at a given indentation depth as detailed in Section II B. Details on the signals measured by the AFM and the Lock-in amplifier are reported in Section II E.

## B. Theory Background and MoNI/ÅI Indentation Curves

During the indentation, for each constant normal force $F_0^z$ the lock-in drives the fixed piezotube oscillation amplitude $\Delta z_{\text{piezo}}$, which is equal to the sum of the cantilever bending $\Delta z_{\text{lever}}$ and the displacement of the tip-sample contact normal to the plane $\Delta z_{\text{indent}}$ as shown in Figure 2(a), so that:

$$\Delta z_{piezo} = \Delta z_{lever} + \Delta z_{indent} \qquad (1)$$

The stiffness of the AFM cantilever and the tip-sample contact can be considered as the connection in series of two springs: the cantilever with stiffness $k_{lev}$ and the tip-sample contact with stiffness $k_{cont}$ (see schematic in Figure 2a). The force required to stretch these two springs in series with a total displacement $\Delta z_{piezo}$ is equal to the normal force variation $\Delta F(F_0^z)$, which depends on the normal force $F_0^z$ and is measured using the Lock-in amplifier during MoNI/ÅI experiments. This experimental configuration allows us to measure the total stiffness $k_{tot}$ at each normal load $F_0^z$, fixed by the feedback loop of the AFM:

$$\frac{\Delta F(F_0^z)}{\Delta z_{piezo}} = k_{tot}(F_0^z) = \left(\frac{1}{k_{lev}} + \frac{1}{k_{cont}(F_0^z)}\right)^{-1} \qquad (2)$$

Therefore, from the measurements through the Lock-in amplifier of $\Delta F(F_0^z)$ for each $F_0^z$ and the knowledge of $\Delta z_{piezo}$ (output signal from the Lock-in amplifier into the piezotube), we can obtain the full $k_{tot}(F_0^z)$ curves as shown in Figure 2(d). Furthermore, since $k_{lev}$ is known, the measurement of $\Delta F(F_0^z)/\Delta z_{piezo}$ at different normal loads $F_0^z$ allows us to acquire the stiffness $k_{cont}(F_0^z)$ as a function of $F_0^z$. If the sample-substrate deforms during indentation, an additional term, $1/k_{sample-substrate}$, needs to be added into equation (2).

The contact stiffness $k_{cont}(F_0^z)$ is by definition equal to:

$$k_{cont}(F_0^z) = \frac{dF_0^z}{dz_{indent}} \qquad (3)$$

where $z_{indent}$ is the indentation depth, which is the maximum normal displacement of the tip-sample contact. By substituting equation (2) in (3) and computing the integral in $F_0^z$, the MoNI/ÅI indentation curve is obtained as

$$z_{indent} - z_{P0} = \int_{F_{P0}}^{F_0^z} \frac{1}{k_{cont}(F_0^{z\prime})} dF_0^{z\prime} = \int_{F_{P0}}^{F_0^z} \left(\frac{1}{k_{tot}(F_0^{z\prime})} - \frac{1}{k_{lev}}\right) dF_0^{z\prime} \qquad (4)$$

where $F_{P0}$ is the pull-out force measured by the AFM when the tip loses contact with the sample's surface at $z_{P0}$. Equations (2) and (4) allow the identification of the effective contact stiffness and computation of the indentation curve for the MoNI/ÅI experiments:

$$z_{indent} - z_{P0} = \int_{F_{P0}}^{F_0^z} \left(\frac{\Delta z_{piezo}}{\Delta F(F_0^{z\prime})} - \frac{1}{k_{lev}}\right) dF_0^{z\prime} \qquad (5)$$

The elastic modulus of the material can be quantified by determining the functional dependence of the contact stiffness $k_{cont}(F_0^z)$ on $F_0^z$. A viable model to study the contact between the tip and the sample is the classical Hertz model for the contact between a sphere and an elastic half space [38, 39]. This model assumes that the contact is frictionless, non-adhesive, and that the area of contact is much smaller than the characteristic radius of the sphere. In addition, the deformation of both the tip and the surface are in the linear elastic regime, and thus fully reversible[1]. Under these assumptions, the force $F_0^z$ can be obtained as function of the indentation depth as

$$F_0^z = \frac{4E^*\sqrt{R}}{3} z_{indent}^{3/2} \qquad (6)$$

and the stiffness of the contact can be determined using equation (3) and (6), so that

$$k_{cont}(F_0^z) = \frac{3}{2}(F_0^z)^{1/3}\left(\frac{4E^*\sqrt{R}}{3}\right)^{2/3} \qquad (7)$$

where $R$ is the radius of the tip and

$$E^* = \left(\frac{1-\nu_{tip}^2}{E_{tip}} + \frac{1-\nu^2}{E}\right)^{-1} \qquad (8)$$

is the contact modulus, where $E_{tip}$ and $\nu_{tip}$ are the elastic modulus and Poisson's ratio of the AFM tip, and $E$ and $\nu$ are the indentation modulus and Poisson's ratio of the sample, respectively.

Notably, while the Hertz model is valid in the case of non-adhesive contact, the Johnson-Kendall-Roberts (JKR) and the Derjaguin-Muller-Toporov (DMT) are popular models for predicting the contact behavior in the presence of adhesive forces [19, 40-42]. In particular, the JKR model is often adopted in the case of contact with compliant solids, whereby the radius $R$ is large and the adhesion forces are also large. More accurate for the case of MoNI/ÅI experiments on stiff solids is the DMT model [19], which instead assumes small $R$ and small long-distance adhesion forces. In the presence of adhesion forces, the modified form of the contact stiffness for the DMT model is

$$k_{DMT}(F_0^z) = \frac{3}{2}(F_0^z + 2\gamma\pi R)^{1/3}\left(\frac{4E^*\sqrt{R}}{3}\right)^{2/3} \qquad (9)$$

where $\gamma$ is the adhesion energy. In the DMT model in equation (9), we have that the pull-out force $F_{P0}$ measured when the tip loses contact with the sample at $z_{P0}$ is equivalent to the resulting adhesion force, so that $F_{P0} = -2\gamma\pi R$. Therefore, the effect of adhesive forces is not neglected in the MoNI/ÅI method, whereby indentation curves are obtained through the corrected "absolute" normal load $F_0^z = F_0^z - F_{P0}$ and "absolute" indentation depth $z_{indent} = z_{indent} - z_{P0}$ in equation (4).

---

[1] Notably, this hypothesis better applies to MoNI/ÅI than to traditional nanoindentation methodologies where plastic deformation of the surface is commonly observed.

To clarify the meaning of the physical quantities discussed in this section, we simulate in Wolfram Mathematica a MoNI/ÅI experiment performed on a graphite sample (indentation modulus $E_\perp$ =30 GPa and $v$=0.2) with a diamond coated tip ($E_{tip}$ =1050 GPa and $v_{tip}$ =0.2) of radius $R$= 100 nm. In this example, the piezotube oscillation $\Delta z_{piezo}$ is computed for amplitude 0.7 Å and frequency of 4 Hz. The oscillation $\Delta z_{piezo}$ as a function of the force $F_0^z$ is displayed in Figure 2(b). When $F_0^z$ goes to zero as the contact between the sample and the tip is lost, $\Delta z_{piezo}$ continues to oscillate in the simulation with the cantilever in air, whereby no elastic forces are applied on the sample and the cantilever. The amplitude and frequency of $\Delta z_{piezo}$ are in fact controlled in the experiment by the Lock-in amplifier, which is independent from the AFM feedback loop that controls $z_0^{AFM}$ and $F_0^z$. In Figure 2(c), we display the force $\Delta F(F_0^z)$ as a function of the force $F_0^z$ for the Hertz model. As expected for the simple Hertzian case, the oscillatory force $\Delta F(F_0^z)$ decreases with $F_0^z$ and eventually goes to zero when $F_0^z = 0$. The value of the effective stiffness computed using equations (7) as a function of $F_0^z$ is displayed in Figure 2(d). The total stiffness computed for $k_{lev}$ = 50 N/m using equation (2) is also displayed in Figure 2(d). The indentation curve obtained by computing the integral in equation (4) are displayed in Figure 2(e). Determination of an accurate indentation curve is the goal of MoNI/ÅI experiments. In what follows, we will detail the different steps necessary to achieve high accuracy in the measurement.

## C. Calibration of Piezotube, Cantilever, and Photodetector

Calibrations of the cantilever spring constant and tip radius, photodetection/deflection of the cantilever, and piezotube oscillation are required before performing the MoNI/ÅI experiments to know: i) the force applied; and ii) the amplitude of the oscillations of the piezotube. These procedures are reported in what follows.

To calibrate the spring constant of the cantilever, we use the Sader method reported in References [43, 44]. Following [44], the spring constant is given as

$$k_{lev} = M_e \, \rho \, bhL \, \omega_{vac}^2 \qquad (10)$$

where ρ is the density of the cantilever, $h$, $b$, and $L$ are the thickness, width and length of the cantilever, respectively, and $\omega_{VAC}$ is the first flexural resonant frequency in vacuum. In addition, following [43], an effective mass $M_e$ = 0.2427 is employed for L/b >5. For cantilevers employed in MoNI/ÅI experiments, we can assume $\rho$ = 2.239 g cm$^{-3}$, which is the nominal density of highly doped silicon. The resonance frequency can be measured in air with the tuning method in non-contact mode (the relation between the resonant frequency in air and in vacuum is discussed in reference [43]). The dimension of the cantilever can be measured by Scanning Electron Microscope (SEM). For example, for a cantilever having thickness $h$ = 4 µm, width $b$ = 25 µm, length $L$ = 123 µm and $\omega_{vac}$ = 2.764 x 10$^6$ s$^{-1}$ (440 kHz) the spring constant is $k_{lev}$ = 53 N/m.

In addition to the measurement of the cantilever spring constant, analysis of the MoNI/ÅI data requires calibration of the AFM tip radius $R$ in equation (7), which can be performed using two different methods. The first method is by SEM imaging, whereby the radius of the tip is directly measured from the images. The second method is the so-called "reference material method" [31]. The MoNI/ÅI measurement is conducted on a well-known isotropic material of which the Young's moduli is known (for example a sapphire crystal) and nonlinear fitting procedure is then employed to fit the $k_{tot}$ versus $F_0^z$ curves while keeping $E^*$ fixed and using $R$ as free fitting parameter (see equation (7)). Results from the two methods can be compared to ensure the accuracy of the measurement.

Calibration of the AFM force-displacement detection system is performed in two steps. In the first step, the voltage associated to the deflection of the cantilever is measured using the photodetector as a function of the displacement of the piezotube $z_{piezo} = z_0^{AFM}$ to determine the sensitivity $W_{photodetector\text{-}piezo}$ (this curve is usually referred as Force-Distance curve in the literature). The second calibration step is required to estimate the sensitivity $S_{voltage\text{-}photodector}$, which relates the voltage associated to the cantilever deflection measured by the photodetector to the voltage applied to the piezotube by the Lock-in amplifier $\Delta V^{Lock-in}(t)/D$. The values of $S_{voltage\text{-}photodetector}$ and $W_{photodetector\text{-}piezo}$ are used to estimate the amplitude of oscillation of the piezotube due to a given voltage applied by the Lock-in amplifier

$$\Delta z_{piezo}(t) = \left(W_{photodetector-piezo} S_{voltage-photodetector}\right) 2\sqrt{2}\, \Delta V^{Lock-in}(t)/D. \qquad (11)$$

Figure 3(a) displays the cantilever deflection measured by the AFM as a function of the displacement of the piezotube $z_{piezo} = z_0^{AFM}$ (Force-Distance curve). The Force-Distance curve is measured using the proprietary software of the AFM microscope (PicoView 1.2). The slope of the Force-Distance curve is used to estimate the sensitivity $W_{photodetector-piezo}$ = 54 nm/V through a linear fitting procedure in Python. The force applied by the cantilever on the surface is computed by multiplying the deflection signal by the cantilever stiffness $k_{lev}$, which is estimated using the Sader method discussed above.

To perform the calibration of $S_{voltage-photodetector}$ a National Instruments GPIB-USB-HS Adapter IEEE 488 Controller is used to interface the Lock-in amplifier with Labview and simultaneously control the frequency and amplitude of the output voltage from the Lock-in into the piezotube ($\Delta V^{Lock-in}(t)/D$ in Figure 1) and acquire the input/output signals from the AFM photodetector (raw and filtered deflections). As a preliminary step in the calibration procedure, the dynamical response of the cantilever in the frequency interval encompassing the Lock-in oscillation frequency, which in this work is 991 Hz, is measured. In Figure 3(b), the signal generated by the photodetector (proportional to the cantilever deflection) measured by the Lock-in is displayed as a function of the frequency used to actuate the piezotube for a fixed value of the applied voltage for free tip vibrations (black curve) and for tip in contact with Silicon Carbide (red curve). The frequency response in air show a peak at approximately 550 Hz, which can be attributed to an internal resonant frequency of the AFM/piezotube, rather than a resonant frequency of the cantilever (first resonant frequency is generally above 200 kHz for cantilevers used in MoNI/ÅI).

The frequency response of the deflection signal in the region encompassing 991 Hz does not show any specific features. This condition is desired for MoNI/ÅI experiments, whereby a small oscillation amplitude is required to improve the resolution of the measurement. The frequency sweep with the tip in hard contact with a Silicon Carbide surface[2] is also displayed in Figure 3(b). Notably, while the resonant frequency can still be identified at 550 Hz, different from the experiment in air only the in-phase component of the frequency response is detected by the Lock-in amplifier. This result is expected, since the quadrature response of the oscillation is almost completely damped when the tip is in contact.

Figure 3(c) displays the cantilever deflection measured by the Lock-in at 991 Hz as a function of the amplitude of the voltage imposed by the Lock-in amplifier on the piezotube when the tip is in very hard contact ($F_0^z = 400$ nN) with a hard surface (Silicon Carbide). As expected, a linear relationship is observed between the applied voltage (piezotube displacement) and the photodetector signal (cantilever bending), with a slope corresponding to a sensitivity $S_{\text{voltage–photodetector}}$ of 1.2 V/V. For completeness, in the inset of Figure 3(c), we report the value of the efficiency of the divider as a function of the dividing factor (defined as $e = S_{\text{voltage-photodetector}}(D=10)/ S_{\text{voltage-photodetector}}(D)$). Notably, a dividing factor above 20 corresponds to a higher $S_{\text{voltage-photodetector}}$, which means that the divider is less efficient for higher dividing factors.

## D. MoNI/ÅI Noise Analysis

The level of AFM noise during MoNI/ÅI measurements can be estimated experimentally by measuring the raw cantilever deflection signal while the tip is kept in contact ($F_0^z = 400$ nN) with a stiff and flat surface (e.g. Silicon Carbide). A periodic voltage signal of 0.14 mV RMS, corresponding to an actual piezotube oscillation of approximately 0.25 Å, is applied to the piezotube at 991 Hz. This signal level is comparable to the one applied during a MoNI/ÅI experiments. The raw deflection signal measured by the AFM controller is acquired using a National Instrument USB-6259 data acquisition board with an acquisition frequency of 40 kHz.

---

[2] Silicon carbide is selected in this experiment for its atomically flat surface.

The raw deflection signal recorded during one of the experiments is displayed in Figure 4(a). Notably, the level of random noise in the raw deflection signal is such that the oscillation at 991 Hz cannot be clearly identified in the response (red oscillatory curve in Figure 4(a)). To analyze the frequency components in the AFM raw deflection, the fast Fourier transform of the raw signal is computed in Figure 4(b). Through the analysis of the raw deflection signal measured by the AFM, we can identify three different components: i) the signal of the cantilever deflection at 991 Hz due to the applied voltage on the piezoelectric element; ii) a signal that is attributed to the power line in two different components at 60 and 120 Hz; iii) the background broadband noise, mainly related to the shooting noise on the photodetector [45].

To isolate the cantilever oscillation from the background signals, the cantilever deflection at 991 Hz is numerically isolated applying a Butterworth bandpass filter in Python (cut-off frequencies: 978 Hz, 1003 Hz). The resulting signal is superimposed to the raw deflection in Figure 4(a) (red dashed curve). As expected, the resulting amplitude of the component at 991 Hz is much smaller than the amplitude of the broadband noise recorded by the AFM. As discussed in Reference [34], similar to our numerical procedure, the Lock-in amplifier performs a digital filtering of the incoming raw deflection signal. The filter band employed by the Lock-in amplifier is as narrow as 0.01 Hz, a bandwidth that we are not able to implement with the numerical Butterworth filter (bandwidth 25 Hz) without cutting a substantial part of our signal. The deflection signal computed through the numerical filtering is compared to the deflection signal obtained through digital filtering by the Lock-in amplifier in Figure 4(c). As expected, we observe that the amplitudes of the two signal are comparable.

For the signal displayed in Figure 4(a), we compute an RMS noise voltage from the raw AFM deflection data of 5 mV (the RMS noise is computed as the square root of the noise signal). An estimation of the deflection in nanometers corresponding to this noise level is obtained by multiplying the voltage signal by the sensitivity $W_{\text{photodetector}-\text{piezo}}$ = 54 nm/V, which results in an equivalent displacement of 0.7 nm. This level of noise is way above the resolution required for the MoNI/ÅI experiment (below 1 Å): this result clearly demonstrates that MoNI/ÅI experiments would not be possible without leveraging the phase sensitive detection of the Lock-in amplifier. The deflection measured by using either the frequency Butterworth filter or directly obtained from the Lock-in amplifier (Figure 4(c)) is ~ 0.25 mV RMS, which would be only the 0.5% of the deflection signal associated to the noise measured when the tip is in contact with a Silicon Carbide surface with a contact force of approximately 400 nN.

To further clarify the role of the noise in MoNI/ÅI experiments, we simulate the effect of noise on the measurement of the contact stiffness $k_{\text{cont}}(F_0^z)$ in equation (7) by adding to the deterministic signals generated by the Lock-in and AFM detection system an additional stochastic noise component [46]. The analysis is performed by superimposing a uniform random white noise of 5 mV RMS, corresponding to a deflection noise of 0.7 nm as measured in our experiments, to the deterministic value of the signal measured by the AFM. In addition, a white random noise level of 1.1 mV RMS is added to the signal of the Lock-in amplifier $\Delta z_{\text{piezo}}$, corresponding to a displacement of the piezotube of 0.02 nm. The contact stiffness in the presence of the noise on the input/output signals is defined as:

$$\tilde{k}_{\text{cont}}^t(F_0^z) = \left( \frac{(\Delta z_{piezo} + \widetilde{\Delta z}_{\text{noise}}^t)}{(\Delta F(F_0^z) + \widetilde{\Delta F}_{\text{noise}}^t(F_0^z))} - \frac{1}{k_{\text{lev}}} \right)^{-1} \tag{12}$$

where the hat superscript indicates noise variables with uniform white noise distribution; the superscript $t$ emphasize the dependence of the uniform random noise on time. The signal to noise ratio (SNR) is computed from equation (12) as the ratio of the expected value of the contact stiffness $k_{\text{cont}}(F_0^z)$ and the standard deviation of the noisy signal $\tilde{k}_{\text{cont}}^t(F_0^z)$ from the expected value $k_{\text{cont}}(F_0^z)$. Thus, the signal to noise ratio is computed as:

$$\text{SNR} = \frac{\frac{1}{N} \sum_{i=1}^{N} k_{\text{cont}(i)}(F_0^z)}{\sqrt{\frac{1}{N} \sum_{i=1}^{N} \left( \tilde{k}_{\text{cont}(i)}^t(F_0^z) - k_{\text{cont}(i)}(F_0^z) \right)^2}} \tag{13}$$

where N is the number of sampling points used to reconstruct the noisy signal. The computation of $\tilde{k}_{\text{cont}}^t(F_0^z)$ and SNR is performed in Wolfram Mathematica. For simplicity, a linear dependence of the force $F_0^z$ on time is adopted in order to compute equations (12) and (13). To ensure repeatability of the simulation a random seed equal to 20 is adopted in the RandomSeed function in Mathematica. The values of $k_{\text{cont}}(F_0^z)$ and $\tilde{k}_{\text{cont}}^t(F_0^z)$ computed as a function of $F_0^z$ are reported in Figure 5 (a-c) for values of the sinusoidal oscillation of the piezotube $\Delta z_{\text{piezo}}$ of 1, 10, and 100 nm at a frequency of 991 Hz on a substrate of graphite (indentation modulus $E_\perp$ =30 GPa and $v$=0.2). The noise level increases with respect to the underlying MoNI/ÅI signal with decreasing oscillation amplitude, and, for oscillation amplitude $\Delta z_{\text{piezo}}$= 1 nm ($\Delta z_{\text{piezo}}$ for MoNI/ÅI is <0.1 nm) in Figure 5(c), it is hard to visually separate the $k_{\text{cont}}(F_0^z)$ signal from the noise. Indentation curves obtained from values of $\tilde{k}_{\text{cont}}^t(F_0^z)$ by computing the integral in equation (5) are displayed in Figure 5(d). While indentation curves obtained for $\Delta z_{\text{piezo}}$ equal to 10 nm and 100 nm give reasonably accurate results for graphite (black solid line is the theoretical expectation from equation (6)), the indentation curve for $\Delta z_{piezo}$= 1 nm does not accurately represent the contact due to the high noise level in the $\tilde{k}_{cont}^t(F_0^z)$ signal.

To establish the dependence of the quality of the $\tilde{k}_{cont}^t(F_0^z)$ signal on the oscillation amplitude, $\Delta z_{piezo}$ is used as a parameter in calculating the Signal-to-Noise ratio (SNR) in equation (13). The analysis is conducted for four substrates with stiffness ranging from 30 GPa to 1000 GPa, namely graphite, zinc oxide, sapphire, and diamond. For clarity of presentation, we chose to display the value of the inverse of the SNR in Figure 6 (a-d), whereby an increase of the inverse of SNR represents an increase of the noise level with respect to the expected signal. Results in Figure 6(a-d) clearly show that the decrease in oscillation amplitude results in a substantial increase in the noise level with respect to the MoNI/ÅI signal. In addition, we observe that for stiffer substrates the noise level increases faster with decreasing $\Delta z_{piezo}$ amplitude. Notably, a SNR$^{-1}$ of approximately 5 (~14 dB) is calculated for diamond at $\Delta z_{piezo}$= 15 nm in Figure 6(d), while a SNR$^{-1}$<1 (<0 dB) is calculated for graphite at the same oscillation amplitude in Figure 6(a). This result demonstrates how a stiffer substrate poses a greater challenge for indentation measurements, and how the high sensitivity of the MoNI/ÅI measurement is beneficial for the characterization of high modulus thin films. In addition, we observe how the SNR$^{-1}$ value sharply increase for $\Delta z_{piezo}$ approaching 1-10 nm depending on the substrate. From our simulation, the expected SNR$^{-1}$ at $\Delta z_{piezo}$ = 0.1 Å in a MoNI/ÅI on diamond is approximately 76 dB, which is estimated through fitting the simulated data with a hyperbolic function (SNR$^{-1}$=65.8/$\Delta z_{piezo}$ for diamond). Therefore, digital filtering of the signal through the Lock-in amplifier (dynamic range 100dB [34]) is absolutely necessary in MoNI/ÅI experiments to effectively isolate the signal from the noise.

To further understand the importance of the digital filters applied through the Lock-in for the sensitivity of the measurement, we can consider the case of a diamond substrate (elastic modulus E=1050 GPa). For this material, when a force $F_0^z = 85$ nN is applied ($W_{photodetector-piezo} = 50$ nm/V) the contact stiffness is equal to 1054 N/m, as computed using equation (7) and (8) for a cantilever with stiffness 170 N/m and tip radius $R$=100 nm [35]. Thus, if we apply an input voltage of 0.1 mV RMS from the Lock-in amplifier during the MoNI/ÅI experiment, a piezo oscillation of 17 pm (0.17 Å, peak to peak) is estimated using equation (11) and an oscillation of the indentation force $\Delta F$ = 2.5 nN is computed using equation (2). This oscillation of the force is used to compute the oscillation of the deflection of the cantilever $\Delta z_{lever}$=15 pm and the oscillation of the indentation depth $\Delta z_{indent}$ = 2 pm (see schematic in Figure 2(a)). In our photodetector, a signal of 17 pm corresponds to a voltage signal of approximately 320 µV, while a signal of 2 pm generates a voltage signal of 44 µV. With a digital lock-in amplifier having a dynamic reserve >

100 dB [34], it is possible to isolate a signal from a noise over $10^5$ times larger than 44 µV, which means a noise of 4.4 V. So when considering the Lab/AFM noise level of 0.7 nm (~ 5 mV) measured in our experiment, we are well within the range of acceptable noise level to perform our measurements. We underline that for lower loads the contact stiffness decreases and therefore the corresponding $\Delta z_{\text{indent}}$ increases, for example for $F_0^z = 10$ nN, $\Delta z_{\text{indent}} = 4$ pm and $\Delta z_{\text{lever}} = 13$ pm.

**E. Lock-In Amplifier Input/Output Signals and Stiffness Measurement**

Figure 7(a) displays the amplitude of the oscillation of the piezotube $\Delta z_{\text{piezo}}$ as a function of the input voltage from the Lock-in amplifier $\Delta V^{\text{Lock-in}}(t)/D$, which is modulated in the range ~0.15-0.8 mV RMS using the Lock-in amplifier. The oscillation amplitude $\Delta z_{\text{piezo}}$ is computed using equation (11) for values of the parameters $W_{\text{photodetector-piezo}} = 54$ nm/V and $S_{\text{voltage-photodetector}} = 1.2$, see Section II C. In the inset, we display the raw signal generated by the amplifier, that is, a sinusoidal voltage input at 991 Hz.

The raw deflection generated by the AFM during a MoNI/ÅI experiment is displayed in Figure 7(b) as a function of time. We can observe an initial jump in the raw deflection located at approximately 5s, which is associated to the initial contact of the tip with the surface. The deflection signal slowly decrease over time while the force applied at the contact is reduced during the MoNI/ÅI experiment. The small bump in the raw deflection signal at 25s is attributed to the initial loss of contact between the tip and surface. Notably, MoNI//ÅI experiments are commonly performed in retraction, meaning the load applied to the tip is progressively reduced until the contact with the surface is loss. This particular procedure is recommended to avoid the effect of the snap-in contact that is characteristic of this class of AFM force measurements.

The raw signal displayed in Figure 7(b) is fed to the Lock-in amplifier to isolate the component at 991 Hz as discussed in Section II D. In Figure 7(c), the component of the cantilever deflection at the fixed frequency 991 Hz measured by the Lock-in amplifier is also displayed as a function of time. The cantilever deflection in Figure 5(c) is used in the data processing to calculate $\Delta F(F_0^z)$ by multiplying this signal by the sensitivity $W_{\text{photodetector-piezo}}$ and the stiffness $k_{\text{lev}}$, see for example the experimental curves in Figure 8(a)-(c). The limits of the integral in equation (5), that is, $F_{P0}$ and $z_{P0}$, are also determined using this curve, by detecting from the deflection signal the position where the tip loses contact with the surface.

The total stiffness, the stiffness of the contact, and the indentation depth are computed from the raw curve in Figure 7(c) using equations (2)-(5) in Section II.B. The mechanical stiffness of the surface is identified by fitting the indentation curve computed with Hertz's equation (6) through a nonlinear fitting procedure in Python, where the modulus $E$ is used as the fitting parameter. Notably, under the assumption of a quasi-isotropic behavior, the modulus $E$ determined by MoNI/ÅI can be compared with the elastic modulus (Young's modulus), as assumed by the Hertz contact model in equations (6)-(8). However, we remark that for indentation experiments performed with very shallow indentation depths (few angstroms) and contact areas of few tenths nanometers square - the tip loading is applied on few thousands atoms – the effect of short range anisotropies, as well as sample topography, can play an important role in determining the mechanical response of the material. Therefore, analysis of the topography of the surface of the sample becomes necessary before performing the experiments, in order to identify suitable regions for indentation experiments.

The uncertainty associated to the estimation of $E$ can be evaluated by propagating the experimental error associated to the variables in equation (6). Following [47], equation (6) can be rewritten in the form

$$E = f\left(V_0^{\text{AFM}}, \Delta V^{\text{Lock-in}}/D, \Delta V^{\text{photod.}}, W_{\text{photod.-piezo}}, S_{\text{voltage-photod.}}, k_{\text{lev}}, R, \nu, E_{\text{tip}}, \nu_{\text{tip}}\right). \quad (14)$$

In order to assess the standard deviation associated to the parameters in equation (14), we compute the Taylor series uncertainty propagation. To this aim, we fix the operating AFM voltages $V_0^{\text{AFM}}, \Delta V^{\text{Lock-in}}/D$ and $\Delta V^{\text{photodetector}}$ and we estimate the standard deviation associated to calibration parameters $W_{\text{photodetector-piezo}}$, $S_{\text{voltage-photodetector}}$, cantilever/tip parameters $k_{\text{lev}}, R$, and material parameters $\nu, E_{\text{tip}}, \nu_{\text{tip}}$. More specifically, the standard deviation associated to one of the parameter (e.g. $R$) is determined by computing the partial derivative of $E$ with respect to the parameter ($R$) and then multiply by the associated uncertainty associated (uncertainty on $R$), see Reference [47].

The uncertainty on the parameters in equation (14) is either measured experimentally or estimated following directions given by the manufacturer of the instruments. The value of the sensitivity $W_{\text{photodetector-piezo}} = 54 \pm 1.35$ is estimated by measuring the slope of the force-distance curve, as discussed in Section II.C. The value of the calibration parameter $S_{\text{voltage-photodetector}} = 1.2 \pm 0.014$ is estimated as reported in Section II.C, whereby the uncertainty is obtained from the tolerance (~1%) on the Z sensitivity of the AFM obtained after calibration with the reference grating (Bruker). The value of $k_{\text{lev}} = 80 \pm 3.2$ is determined from the manufacturer (Nanosensors) with an uncertainty of 4%, which is assumed from uncertainty level for AFM cantilever estimated in [47]. The value of the tip radius $R = 190\pm10$ nm is identified with a margin of uncertainty using the reference material method (sapphire is adopted as the calibration material). This value of the tip radius is within the range declared by the supplier (100-200 nm) and is verified using SEM imaging. Finally, the values of $\nu = 0.2, E_{\text{tip}} = 1050, \nu_{\text{tip}} = 0.2$ are adopted for the sample and the tip, with an estimated uncertainty of 5%.

The resulting uncertainty on the estimation of $E$ is strongly dependent on the absolute value of the stiffness of the contact, whereby a decrease of the contact force results in a decrease of the overall uncertainty. For example, for a material with stiffness $E = 400$ GPa, we obtain a standard deviation of 75 GPa (~19%). The standard deviation is reduced to 6.5 GPa (~7%) at 100 GPa (we remark that this analysis is conducted based on conservative assumptions on the underlying uncertainty of the parameters). At $E = 100$ GPa, the standard deviation associated to the uncertainty on the sensitivity $S_{\text{voltage-photodetector}}$ is 3.9 GPa, which corresponds to the 36% of the overall variance of the measurement. The sensitivity $S_{\text{voltage-photodetector}}$ is therefore the dominant calibration parameter in determining the accuracy of the force measurement, as already discussed in Reference [47]. The remaining variance of the measurement is mainly distributed between the uncertainty associated to the estimation of $k_{\text{lev}}$ (31%) and $R$ (24%), which proves that a careful estimation of the tip properties is necessary to obtain accurate measurements using MoNI/ÅI.

## III. MONI/ÅI APPLIED TO ULTRA-THIN AND ULTRA-HARD FILMS

### A. Application to Stiff and Thin Films

We use MoNI/ÅI to identify the mechanical properties, namely the indentation elastic modulus, of an epitaxial thin film of CVD diamond (2.2 µm on Silicon), a sapphire substrate (Al$_2$O$_3$, polycrystalline), and a zinc oxide crystal (ZnO). These materials are selected as references and testing systems for MoNI/ÅI experiments due to their well-known mechanical properties and chemical stability. Results of the MoNI/ÅI experiments performed on the three reference materials are reported in Figure 8. The indentation is performed in 5-6 different positions on each sample to give an estimation of the uncertainty in the measurement. The total stiffness and contact stiffness obtained from MoNI/ÅI experiments are displayed in Figure 8(a)-(f), wherein the shaded area represent one standard deviation from the mean. We underline that while $k_{\text{tot}}(F_0^z)$ is directly measured as described before, $k_{\text{cont}}(F_0^z)$ is obtained using equation (2). For all the experiments, the same AFM probe has been employed with a cantilever stiffness $k_{\text{lev}}$ of 80 N/m of a sensitivity $W_n$ of 54 nm/V. The probe tip is coated with diamond ($E_{\text{tip}} = 1050$ GPa, $v_{\text{tip}} = 0.2$). During this experiments, we used $\Delta V^{\text{Lock-in}} = 4$ mV RMS, $D = 20$ (e ~ 1), so that the amplitude of oscillation is $\Delta z_{\text{piezo}}$=0.34 Å.

Figure 8(a) and (d) displays the total and contact stiffness obtained for the CVD diamond film on Silicon. The total stiffness measured by MoNI/ÅI is close to 80 N/m, which indicates an extremely high stiffness of the surface[3]. A contact stiffness ranging between 2000 and 3000 N/m is measured when the tip is in hard contact with the sample. Contact stiffness substantially decreases with decreasing contact force below 20 nN, and goes to zero as the contact between the tip and the sample is lost. Notably, the error in the estimation of the contact stiffness sharply increases below 20 nN, which might likely be attributed to the uneven surface of the CVD sample, which can affect the effective contact area when the force applied by the tip is reduced, even if the indentation region is carefully selected from the topography in Figure 9.

---

[3] This result can be explained by looking at the schematic in Figure 3(a). When the stiffness of the contact spring approaches infinity, the stiffness of the series connection become equivalent to $k_{\text{lev}}$.

Figure 8(b) and (e) displays the total stiffness and contact stiffness of the sapphire substrate. The value of the total stiffness also in this case is close to the stiffness of the cantilever (80 N/m), which indicates the high stiffness of the surface. The values of the contact stiffness for the sapphire range between 1500 and 2000 N/m. Results obtained for the ZnO crystal are reported in Figure 8(c) and (f). In this case, the total stiffness is of the order of 70 N/m due to the lower stiffness of the ZnO compared to the CVD diamond and the sapphire. The lower total stiffness is related to the substantially softer contact with $k_{\text{cont}}$ below 1000 N/m. The standard deviation associated to measurements performed on ZnO and sapphire samples is lower than the standard deviation obtained for CVD diamond. This can be attributed to a smoother surface, as observed in the topography in Figure 9, as well as to the lower stiffness of the materials that is associated with a stronger MoNI/ÅI signal $\Delta F(F_0^z)$ with respect to the instrumentation noise, as discussed in Section II.D.

The indentation curves in Figure 8(h)-(i) are computed from the contact stiffness in Figure 8(d)-(e) using equation (5). Qualitative analysis of the indentation curves shows how the stiffness of the CVD diamond film is higher than the stiffness of sapphire and ZnO (steeper curves). We also observe that the standard deviation on the measurement of the indentation curve for CVD diamond is higher than the standard deviation for sapphire and ZnO, which is well in line with our estimation of the noise in the measurement of the contact stiffness, as shown in Figure 6. The quantitative analysis of the indentation data is performed by fitting the experimental curves with equations (6)-(8), as already discussed in this Section II. Results of the fitting procedure for indentation curves in Figure 8 are reported in Figure 10(a), together with the values of the indentation elastic moduli for the three materials. Equation (5) is used to identify the stiffness of the three materials. Results of this procedure give a prediction of the indentation moduli of 117±12, 416±45, and 1139±296 GPa for ZnO, sapphire, and CVD diamond, respectively. These values are in excellent agreement with values of the elastic moduli expected for these materials, which confirm that MoNI/ÅI is able to give a quantitative estimation of the elastic moduli of very stiff materials in a non-destructive way with high spatial resolution. Normalized distributions of indentation moduli computed for aggregated data for 15-30 experiments obtained for each material are displayed in Figure 10(b). For completeness, the distribution of the indentation modulus for a $SiO_2$ substrate ($E$~60 GPa) is also reported, in order to extend the range of stiffness of materials tested. Notably, narrower distributions are obtained for SiO$_2$ and ZnO (plot in log scale) with respect to sapphire and CVD diamond. This result is well in line with results previously outlined in Section II.E, whereby higher contact stiffness is associated with higher uncertainty on the estimation of the contact modulus. Distributions obtained for both CVD diamond and sapphire are sufficiently narrow to clearly identify the indentation modulus, further demonstrating the application of MoNI/ÅI to direct in-situ measurement of the mechanical properties of ultra-stiff materials.

## B. Application to 2D Materials for Transverse Elasticity and Carbon Nanofilaments

Differently from traditional nanoindentation, MoNI/ÅI can be used to accurately characterize the mechanical properties of stiff surfaces on extremely small scales (sub-Å). MoNI/ÅI has been successfully applied to the mechanical characterization of two-dimensional (2D) materials. The extremely small indentation depth utilized during MoNI/ÅI has enabled for the first time the study of the transverse elasticity of few layer thick graphene films [19]. In particular, indentation depths in the order of few tenths of an Angstrom have been employed to probe the interlayer stiffness of epitaxial graphene and epitaxial graphene oxide, and, more recently, to demonstrate the ultra-high stiffness of 2-layer epitaxial graphene [35, 36]. These experiments showed that Van der Waals interactions between different layers play an important role in determining the stiffness of the material when a force is applied perpendicular to the principal planes of graphene.

As an example of the potential of MoNI/ÅI in 2D materials applications, Figure 11(a) displays an indentation curve of 10-layer epitaxial graphene (EG) grown (supported) on the Carbon polar face (000-1) of Silicon Carbide (SiC). Several measurements provided for 10-layer epitaxial graphene an indentation modulus perpendicular to the planes equal to $E_\perp = (36 \pm 3)$ GPa, the same as that of graphite [48]. This is not surprising because graphene can be mechanically regarded as a "thinner version" of graphite, the interlayer van der Waals property should not differ significantly. Notably, the indentation depth in MoNI/ÅI experiments (down to 0.3 Å) is smaller than the interlayer distance between graphene planes (approximately 3.4 Å). Since contribution of the in plane stiffness can be proved to be negligible at these small indentation depths, the experiments allowed to identify the transverse stiffness of the 2D material.

MoNI/ÅI has also been employed for the characterization of the interlayer mechanical properties of epitaxial graphene oxide and graphene oxide [19], and in particular to investigate the effect of water intercalation between the layers. Figure 11(b) displays the MoNI/ÅI indentation curves obtained at 25% relative humidity for epitaxial graphene oxide and graphene oxide flakes deposited on a Silicon wafer. Notably, the effect of water intercalation in the porous structure of graphene oxide results in a much higher modulus (~35 GPa) than the modulus measured for epitaxial graphene oxide (~22 GPa), whereby water intercalation is minimal in epitaxial graphene oxide [19].

MoNI/ÅI measurements have been conducted in parallel with topographic AFM imaging to investigate the mechanical stiffness of nanostructures, such as Q-carbon filaments [17]. This is a characteristic feature of MoNI/ÅI, whereby the AFM can be employed to image the surface (in tapping mode) and perform the elasticity measurement (in contact). In Figure 12(a), we display the topography of the cross section of a Q-carbon filament reconstructed from AFM data, as well as a larger scan (25 μm$^2$) displaying few interconnecting Q-carbon filaments. We also report the positions where the MoNI/ÅI experiments are conducted. In Figure 12(b), we display the indentation curves. Indentation data show a substantial increase in the stiffness of the Q-carbon filament (red lines) compared to the surrounding diamond-like-carbon material (black lines) [17]. Notably, the small lateral width (~200 nm) and depth (~30 nm) of the filament make the measurement of the mechanical properties of the nanostructure with classical nanoindentation techniques extremely difficult. The indentation area of classical nanoindenters will likely be larger than the filament and the resulting stiffness measurement an average of the mechanical stiffness of the filament and the surrounding diamond-like-carbon material. In this case, MoNI/ÅI is a viable solution for mechanical measurement of the nanostructure, whereby an increased spatial resolution is required.

## IV. CONCLUSION

In this report, we have presented a novel methodology, MoNI/ÅI, for non-destructive sub-Å-depth indentation measurements of ultra-stiff and ultra-thin films, as well as two-dimensional materials for in-situ elasticity measurements with nanoscale topographical imaging. During conventional nanoindentation measurements, the indentation depths are usually larger than 10 nm, which hinders the ability to study ultra-thin films (< 10 nm) and supported atomically thin 2D materials. Differently from traditional nano-indentation methods, MoNI/ÅI employs extremely small amplitude oscillations (<< 1 Å), higher frequency, and force detection systems based on laser detection AFM methods. We show that this methodology can be easily implemented with a commercial AFM and we demonstrate that in presence of a standard laboratory noise floor, the signal to noise ratio of MoNI/ÅI is such that a dynamic range of 80 dB — achievable with commercial Lock-in amplifiers— is sufficient to observe superior indentation curves, with indentation depths as small as 0.3 Å, resolution in indentation < 0.05 Å and in normal load < 0.5 nN. We also prove that MoNI/ÅI is a powerful tool to measure the indentation moduli of ultra-stiff ultra-thin films with much higher spatial resolution (both vertically and horizontally) and smaller indentation depths than other indentation methods. Because of its simplicity, and its implementation with commercial equipment,

MoNI/ÅI has the potential for a broad applicability in studying the elasticity of ultra-thin films (< 10 nm) for semiconductor and energy applications, nano-structures, and atomically thin two-dimensional materials (~ 0.5 - 5 nm).

## ACKNOWLEDGMENTS

This research was supported by the Office of Basic Energy Sciences of the US Department of Energy (grant no. DE-SC0018924). The Authors thank Professor Angelo Bongiorno and Dr. Tengfei Cao for inspiring discussion. The Authors thank Francesco Lavini for careful review of this manuscript.

## ADDITIONAL INFORMATION

The authors declare no competing financial and/or non-financial interests in relation to the work described.

## AUTHOR CONTRIBUTIONS

F.C. and Y. G. performed nanomechanics experiments and data analysis. F.C. carried out the experimental and numerical analysis of the nanoindentation process and noise analysis. E.R. conceived and designed the experiments and analyzed the data. F.C. and E.R. contributed to writing the manuscript.

## REFERENCES

1. Li, X.D., and Bhushan, B., *A review of nanoindentation continuous stiffness measurement technique and its applications.* Materials Characterization, 2002. **48**(1): p. 11-36.
2. Oliver, W.C., and Pharr, G. M., *An Improved Technique for Determining Hardness and Elastic-Modulus Using Load and Displacement Sensing Indentation Experiments.* Journal of Materials Research, 1992. **7**(6): p. 1564-1583.
3. Kumar, K.S., Van Swygenhoven, H., and Suresh, S., *Mechanical behavior of nanocrystalline metals and alloys.* Acta Materialia, 2003. **51**(19): p. 5743-5774.
4. Nix, W.D., *Elastic and plastic properties of thin films on substrates: nanoindentation techniques.* Materials Science and Engineering a-Structural Materials Properties Microstructure and Processing, 1997. **234**: p. 37-44.
5. Rho, J.Y., Tsui, T. Y., and Pharr, G. M., *Elastic properties of human cortical and trabecular lamellar bone measured by nanoindentation.* Biomaterials, 1997. **18**(20): p. 1325-30.
6. Roos, W.H., and Wuite, G. L., *Nanoindentation Studies Reveal Material Properties of Viruses.* Advanced Materials, 2009. **21**(10-11): p. 1187-1192.

7.  VanLandingham, M.R., Villarrubia, J. S., Guthrie, W. F., and Meyers, G. F., *Nanoindentation of polymers: An overview.* Macromolecular Symposia, 2001. **167**: p. 15-43.

8.  Gallego-Gomez, F., Morales-Florez, V., Blanco, A., de la Rosa-Fox, N., Lopez, C., *Water-dependent micromechanical and rheological properties of silica colloidal crystals studied by nanoindentation.* Nano Letters, 2012. **12**(9): p. 4920-4.

9.  Hay, J., P. Agee, and E. Herbert, *Continuous Stiffness Measurement during Instrumented Indentation Testing.* Experimental Techniques, 2010. **34**(3): p. 86-94.

10. Nowakowski, B.K., Smith, D. T., Smith, S. T., Correa, L. F., Cook, R. F., *Development of a precision nanoindentation platform.* Rev Sci Instrum, 2013. **84**(7): p. 075110.

11. Nohava, J., N.X. Randall, and N. Conte, *Novel ultra nanoindentation method with extremely low thermal drift: Principle and experimental results.* Journal of Materials Research, 2009. **24**(3): p. 873-882.

12. Guillonneau, G., Kermouche, G., Bec, S., Loubet, J. L., *Extraction of Mechanical Properties with Second Harmonic Detection for Dynamic Nanoindentation Testing.* Experimental Mechanics, 2012. **52**(7): p. 933-944.

13. Guillonneau, G., Kermouche, G., Bec, S., Loubet, J. L., *Determination of mechanical properties by nanoindentation independently of indentation depth measurement.* Journal of Materials Research, 2012. **27**(19): p. 2551-2560.

14. Oyen, M.L. and R.F. Cook, *A practical guide for analysis of nanoindentation data.* J Mech Behav Biomed Mater, 2009. **2**(4): p. 396-407.

15. Dou, L., Wong, A. B., Yu, Y., Lai, M., Kornienko, N., Eaton, S. W., Fu, A., Bischak, C. G., Ma, J., Ding, T., Ginsberg, N. S., Wang, L. W., Alivisatos, A. P., Yang, P., *Atomically thin two-dimensional organic-inorganic hybrid perovskites.* Science, 2015. **349**(6255): p. 1518-21.

16. Chakraborty, C., Kinnischtzke, L., Goodfellow, K. M., Beams, R., and Vamivakas, A. N., *Voltage-controlled quantum light from an atomically thin semiconductor.* Nature Nanotechnol, 2015. **10**(6): p. 507-11.

17. Song, J., Wang, X., Riedo, E., Wang, Z. L., *Elastic property of vertically aligned nanowires.* Nano Lett, 2005. **5**(10): p. 1954-8.

18. Lee, C.H., Lee, G. H., van der Zande, A. M., Chen, W., Li, Y., Han, M., Cui, X., Arefe, G., Nuckolls, C., Heinz, T. F., Guo, J., Hone, J., and Kim, P., *Atomically thin p-n junctions with van der Waals heterointerfaces.* Nat Nanotechnol, 2014. **9**(9): p. 676-81.

19. Gao, Y., Kim, S., Zhou, S., Chiu, H. C., Nelias, D., Berger, C., de Heer, W., Polloni, L., Sordan, R., Bongiorno, A., Riedo, E., *Elastic coupling between layers in two-dimensional materials.* Nature Materials, 2015. **14**(7): p. 714-720.

20. Falin, A., Cai, Q., Santos, E. J. G., Scullion, D., Qian, D., Zhang, R., Yang, Z., Huang, S., Watanabe, K., Taniguchi, T., Barnett, M. R., Chen, Y., Ruoff, R. S., Li, L. H., *Mechanical properties of atomically thin boron nitride and the role of interlayer interactions.* Nat Commun, 2017. **8**: p. 15815.

21. Wei, Z., Wang, D., Kim, S., Kim, S. Y., Hu, Y., Yakes, M. K., Laracuente, A. R., Dai, Z., Marder, S. R., Berger, C., King, W. P., de Heer, W. A., Sheehan, P. E., and Riedo, E., *Nanoscale tunable reduction of graphene oxide for graphene electronics.* Science, 2010. **328**(5984): p. 1373-6.

22. Menges, F., Riel, H., Stemmer, A., Dimitrakopoulos, C., and Gotsmann, B., *Thermal Transport into Graphene through Nanoscopic Contacts.* Physical Review Letters, 2013. **111**(20).

23. Meyer, E., Heinzelmann, H., Grutter, P., Jung, T., Weisskopf, T., Hidber, H. R., Lapka, R., Rudin, H., Guntherodt, H. J., *Comparative-Study of Lithium-Fluoride and Graphite by Atomic Force Microscopy (Afm).* Journal of Microscopy-Oxford, 1988. **152**: p. 269-280.

24. Butt, H.J., B. Cappella, and M. Kappl, *Force measurements with the atomic force microscope: Technique, interpretation and applications.* Surface Science Reports, 2005. **59**(1-6): p. 1-152.

25. Cohen, S.R. and E. Kalfon-Cohen, *Dynamic nanoindentation by instrumented nanoindentation and force microscopy: a comparative review.* Beilstein J Nanotechnol, 2013. **4**: p. 815-33.

26. Solares, S.D., *Nanoscale effects in the characterization of viscoelastic materials with atomic force microscopy: coupling of a quasi-three-dimensional standard linear solid model with in-plane surface interactions.* Beilstein J Nanotechnol, 2016. **7**: p. 554-71.

27. Killgore, J.P., Kelly, J. Y., Stafford, C. M., Fasolka, M. J., Hurley, D. C., *Quantitative subsurface contact resonance force microscopy of model polymer nanocomposites.* Nanotechnology, 2011. **22**(17): p. 175706.

28. Killgore, J.P., Yablon, D. G., Tsou, A. H., Gannepalli, A., Yuya, P. A., Turner, J. A., Proksch, R., Hurley, D. C., *Viscoelastic property mapping with contact resonance force microscopy.* Langmuir, 2011. **27**(23): p. 13983-7.

29. Maivald, P., Butt, H. J., Gould, S. A. C., Prater, C. B., Drake, B., Gurley, J. A., Elings, V. B., Hansma, P. K., *Using force modulation to image surface elasticities with the atomic force microscope.* Nanotechnology, 1991. **2**: p. 103-106.

30. Hurley, D.C., Shen, K., Jennett, N. M., Turner, J. A., *Atomic force acoustic microscopy methods to determine thin-film elastic properties.* Journal of Applied Physics, 2003. **94**(4): p. 2347-2354.

31. Kocun, M., Labuda, A., Meinhold, W., Revenko, I., Proksch, R., *Fast, High Resolution, and Wide Modulus Range Nanomechanical Mapping with Bimodal Tapping Mode.* ACS Nano, 2017. **11**(10): p. 10097-10105.
32. Aureli, M., Ahsan, S. N., Shihab, R. H., Tung, R. C., *Plate geometries for contact resonance atomic force microscopy: Modeling, optimization, and verification.* Journal of Applied Physics, 2018. **124**(1).
33. Herruzo, E.T., A.P. Perrino, and R. Garcia, *Fast nanomechanical spectroscopy of soft matter.* Nat Commun, 2014. **5**: p. 3126.
34. Stanford Research Systems, *MODEL SR830 DSP Lock-In Amplifier*. 2001.
35. Gao, Y., Cao, T., Cellini, F., Berger, C., de Heer, W., Tosatti, E., Bongiorno, A., and Riedo, E., *Ultrahard carbon film from epitaxial two-layer graphene.* Nature Nanotechnology, 2017. **13**(2): p. 133-138.
36. Cellini, F., Lavini, F., Cao, T., de Heer, W., Berger, C., Bongiorno, A., Riedo, E., *Epitaxial two-layer graphene under pressure: Diamene stiffer than Diamond.* FlatChem, 2018. **10**: p. 8-13.
37. Narayan J., G.S., Bhaumik A., Sachan R., Cellini F., Riedo E., *Q-carbon harder than diamond.* MRS Communications, 2018.
38. Willis, J.R., *Hertzian contact of anisotropic bodies.* Journal of the Mechanics and Physics of Solids, 1966. **14**: p. 163-176.
39. Field, J.S. and M.V. Swain, *A Simple Predictive Model for Spherical Indentation.* Journal of Materials Research, 1993. **8**(2): p. 297-306.
40. Barthel, E., *Adhesive elastic contacts: JKR and more.* Journal of Physics D: Applied Physics, 2008. **41**: p. 163001.
41. Carpick, R.W., D.F. Ogletree, and M. Salmeron, *A General Equation for Fitting Contact Area and Friction vs Load Measurements.* J Colloid Interface Sci, 1999. **211**(2): p. 395-400.
42. Ebenstein, D.M. and K.J. Wahl, *A comparison of JKR-based methods to analyze quasi-static and dynamic indentation force curves.* J Colloid Interface Sci, 2006. **298**(2): p. 652-62.
43. Sader, J.E., Larson, I., Mulvaney, P., and White, L. R., *Method for the calibration of atomic force microscope cantilevers.* Review of Scientific Instruments, 1995. **66**(7).
44. Sader, J.E., J.W.N. Chon, and P. Mulvaney, *Calibration of rectangular atomic force microscope cantilevers.* Review of Scientific Instruments, 1999. **70**(10).
45. Voigtländer, B., *Scanning probe microscopy*. 2015, Berlin: Springer

46. Labuda, A., Lysy, M., Paul, W., Miyahara, Y., Grutter, P., Bennewitz, R., Sutton, M., *Stochastic noise in atomic force microscopy.* Phys Rev E Stat Nonlin Soft Matter Phys, 2012. **86**(3 Pt 1): p. 031104.
47. Wagner, R., Moon, R., Pratt, J., Shaw, G., Raman, A., *Uncertainty quantification in nanomechanical measurements using the atomic force microscope.* Nanotechnology, 2011. **22**(45).
48. Kelly, B.T., *Physics of graphite.* Applied Science, London, 1981.

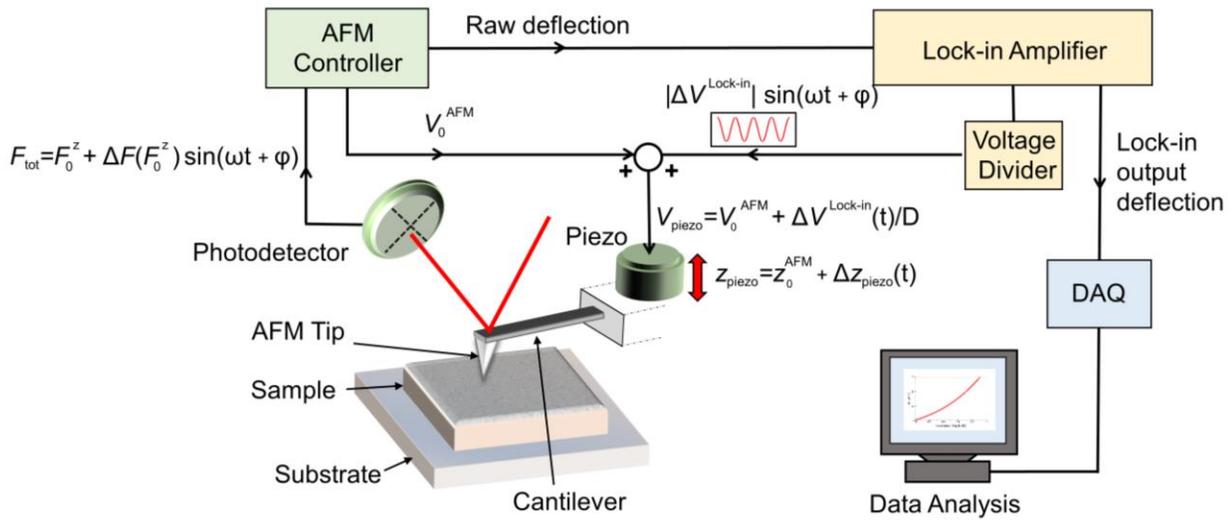

**Figure 1. MoNI/ÅI experiment schematic.** A schematic representation of the setup for MoNI/ÅI. The arrows indicate the input/output directions of the signals from the different devices.

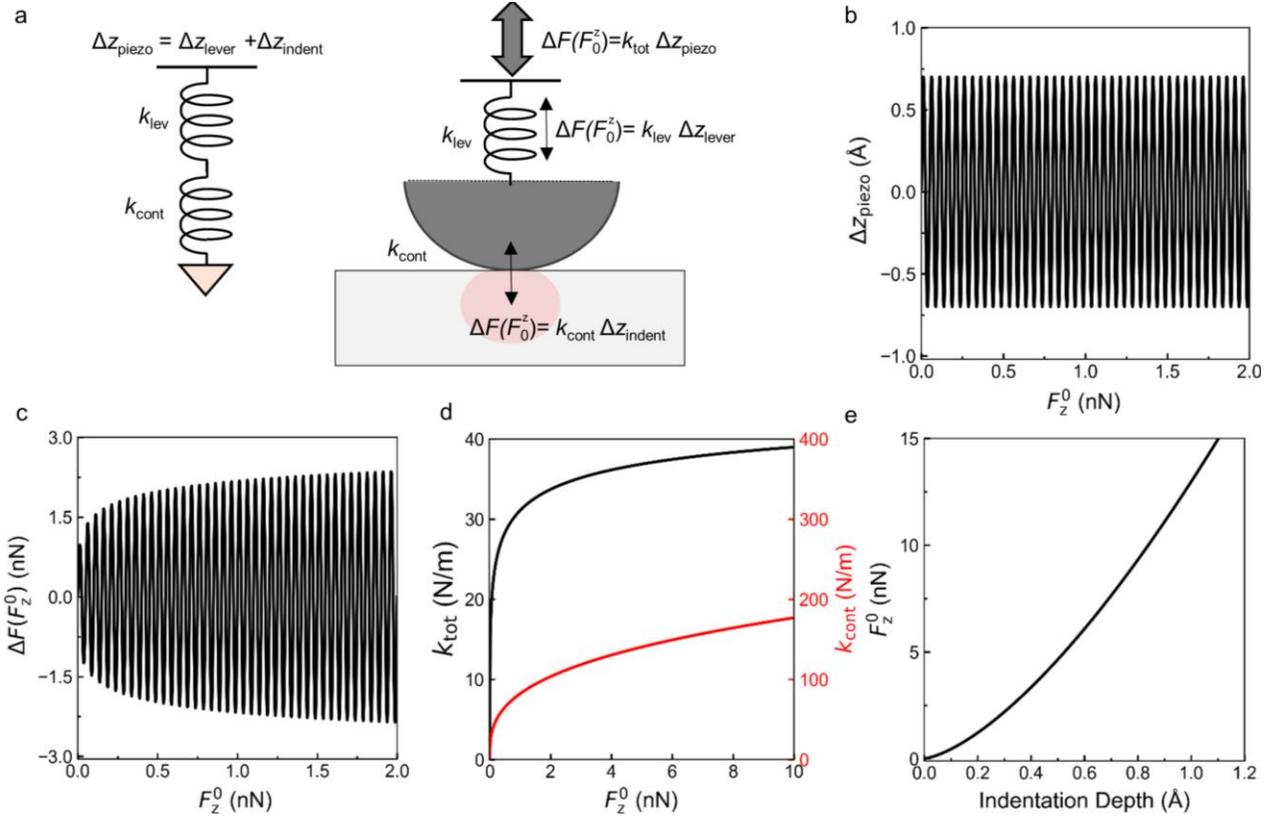

**Figure 2. Simulations of MoNI/ÅI indentation.** (a) Schematic of the contact problem: the red shaded area represents the region of the sample contributing to the effective contact stiffness. (b) Simulated piezotube oscillation $\Delta z_{\text{piezo}}$ computed as a function of the applied force $F_0^z$. Notably, the oscillation of the piezotube $\Delta z_{\text{piezo}}$ in the experiments is imposed by the Lock-in amplifier and it is independent from the displacement $z_0^{\text{AFM}}$, which is controlled through the AFM together with $F_0^z$. (c) Simulated resulting force $\Delta F(F_0^z)$ computed as a function of the applied force $F_0^z$. (d) Simulated total stiffness $k_{\text{tot}}$ (black line) and contact stiffness $k_{\text{cont}}$ (red line) computed as a function of $F_0^z$. (e) Indentation curve computed using equation (4) from the contact stiffness in (d).

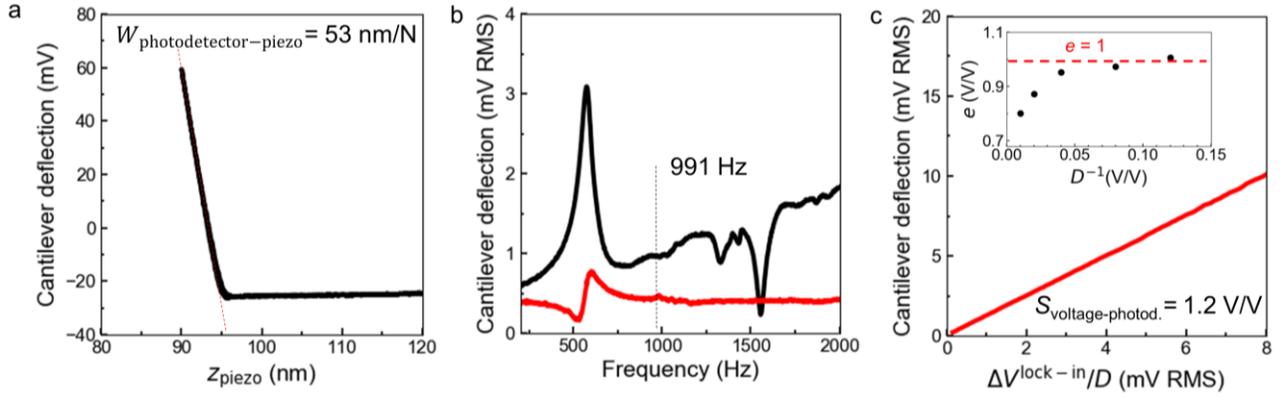

**Figure 3. Experimental calibration curves.** (a) Experimental force distance curve for the cantilever used in the experiments reported in the Section III.A. The sensitivity $W_{\text{photodetector-piezo}}$ is obtained from the slope of the loading curve. (b) Frequency response of the cantilever deflection in the range 200-2000 Hz. Black line is the amplitude of oscillation in air, Red line is the amplitude of oscillation in contact on SiC ($F_0^z = 400\ nN$). (c) Amplitude of oscillation at 991 Hz with the cantilever in contact on SiC ($F_0^z = 400\ nN$) as a function of the lock-in voltage input to the piezo. $S_{\text{voltage-photodetector}}$ is the non-dimensional sensitivity.

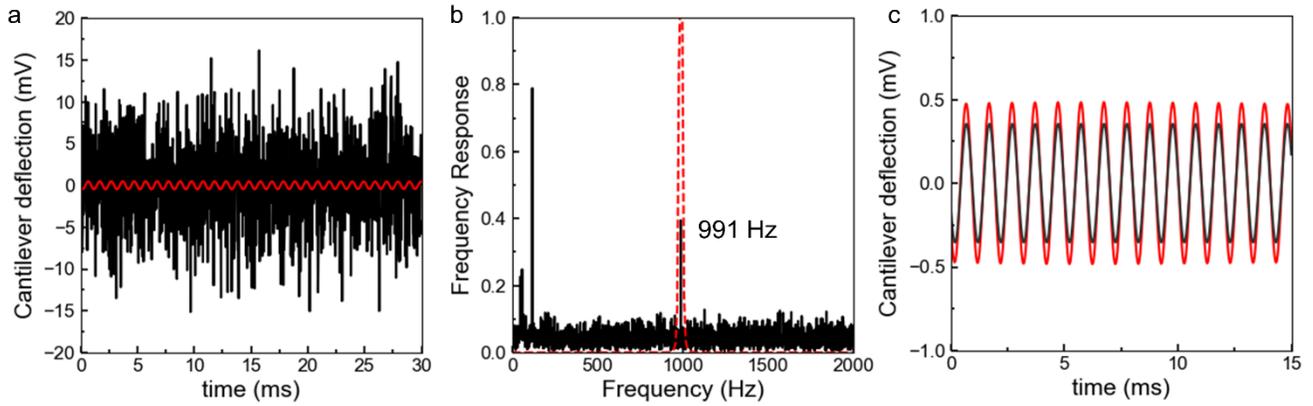

**Figure 4. Experimental noise measurement.** (a) Experimental raw deflection signal recorded by the AFM controller over 30 ms (black line) and the signal component at 991 Hz isolated using the bandpass filter depicted in (b). (b) Fast Fourier transform computed using Python of the raw deflection signal in (a) (black line) and Butterworth filter adopted to isolate the component at the fixed Lock-in frequency 991 Hz (red dashed line). (c) Displacement signal at 991 Hz detected using the numerical filter (red line) and the digital filter of the Lock-in amplifier (dark gray line).

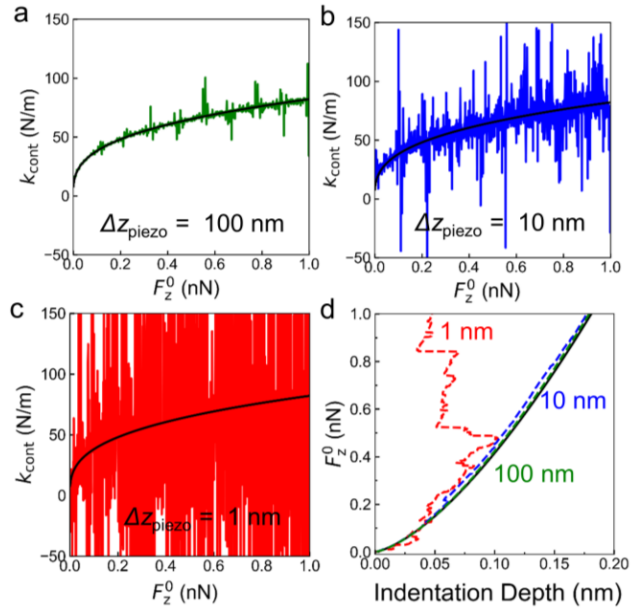

**Figure 5. Simulated $k_{\text{cont}}$ in presence of noise without frequency specific filtering.** (a) Simulated contact stiffness $k_{\text{cont}}$ of a diamond tip (1050 GPa) on Graphite (30 GPa) computed as a function of the applied force $F_0^z$ for reference (solid black line from equation (6)) and noisy signal in equation (12) (solid green line) computed for piezotube oscillation $\Delta z_{\text{piezo}} = 100$ nm. (b) $k_{\text{cont}}$ as a function of $F_0^z$ for reference (solid black line) and noisy signal (solid red line) computed for $\Delta z_{\text{piezo}} = 10$ nm. (c) $k_{\text{cont}}$ as a function of $F_0^z$ for reference (solid black line) and noisy signal (solid red line) computed for $\Delta z_{\text{piezo}} = 1$ nm. (d) Indentation curves computed using equation (5) for values of the contact stiffness displayed in (a-c) for the reference signal (black solid line), and noisy signal with $\Delta z_{\text{piezo}} = 100$ nm (green dashed line), oscillation $\Delta z_{\text{piezo}} = 10$ nm (blue dashed line), and $\Delta z_{\text{piezo}} = 1$ nm (red dashed line).

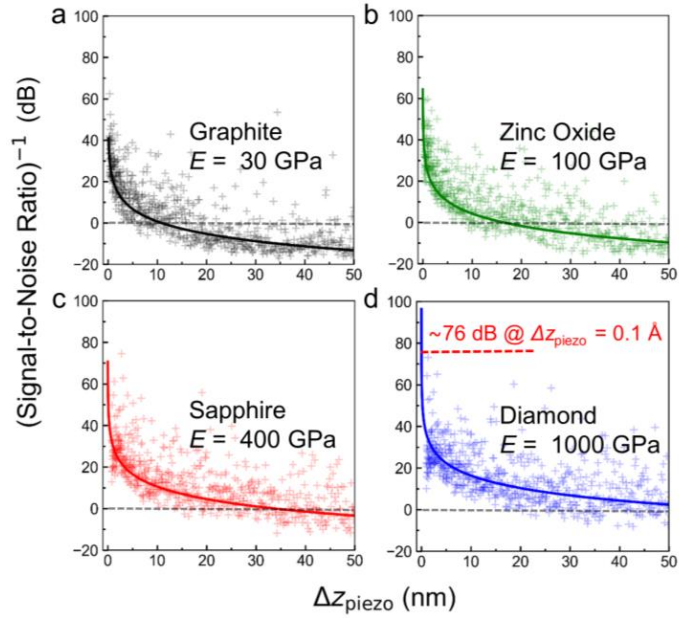

**Figure 6. Inverse of the Signal to noise ratio of $k_{cont}$ as a function of $\Delta z_{piezo}$ for different substrates without frequency specific filtering.** (a) Inverse of the Signal-to-Noise ratio computed using equation (13) as a function of the oscillation amplitude $\Delta z_{piezo}$. Numerical indentation experiments are performed in Mathematica (white noise, uniform distribution, RandomSeed=20) for Hertzian contact between a diamond tip (E=1050 GPa) and Graphite. Solid line is the fitting with a hyperbolic function $SNR^{-1}=10.9/\Delta z_{piezo}$. (b) Inverse of the Signal-to-Noise ratio computed for zinc oxide (100 GPa) as a function of $\Delta z_{piezo}$. Solid line is the fitting $SNR^{-1}=16.4/\Delta z_{piezo}$. (c) Inverse of the Signal-to-Noise ratio computed for sapphire (400 GPa) as a function of $\Delta z_{piezo}$. Solid line is the fitting $SNR^{-1}=33.8/\Delta z_{piezo}$. (d) Inverse of the Signal-to-Noise ratio computed for diamond (1000 GPa) as a function of $\Delta z_{piezo}$. Solid line is the fitting $SNR^{-1}=65.8/\Delta z_{piezo}$.

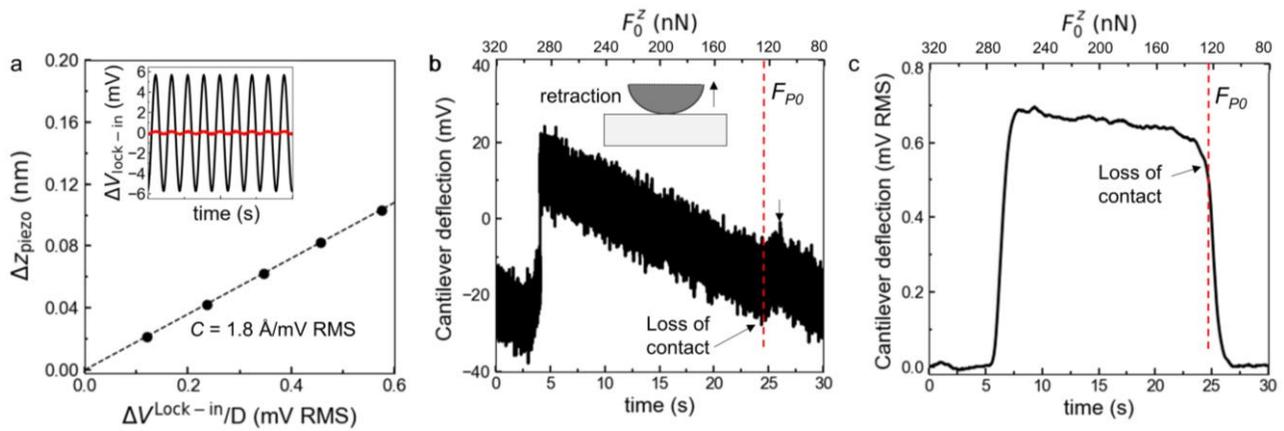

**Figure 7. Experimental MoNI/ÅI signals.** (a) Experimental amplitude of the oscillation as a function of the voltage fed to the piezoelectric element using the Lock-in amplifier. In the inset, the voltage generated by the Lock-in and the corresponding voltage signal after the divider in Figure (1) (experimental values). (b) Raw deflection recorded by the AFM during a MoNI/ÅI experiment measured during retraction of the tip from the surface. (c) Deflection measured by the Lock-in through phase-sensitive detection at 991 Hz from the raw deflection signal in (b).

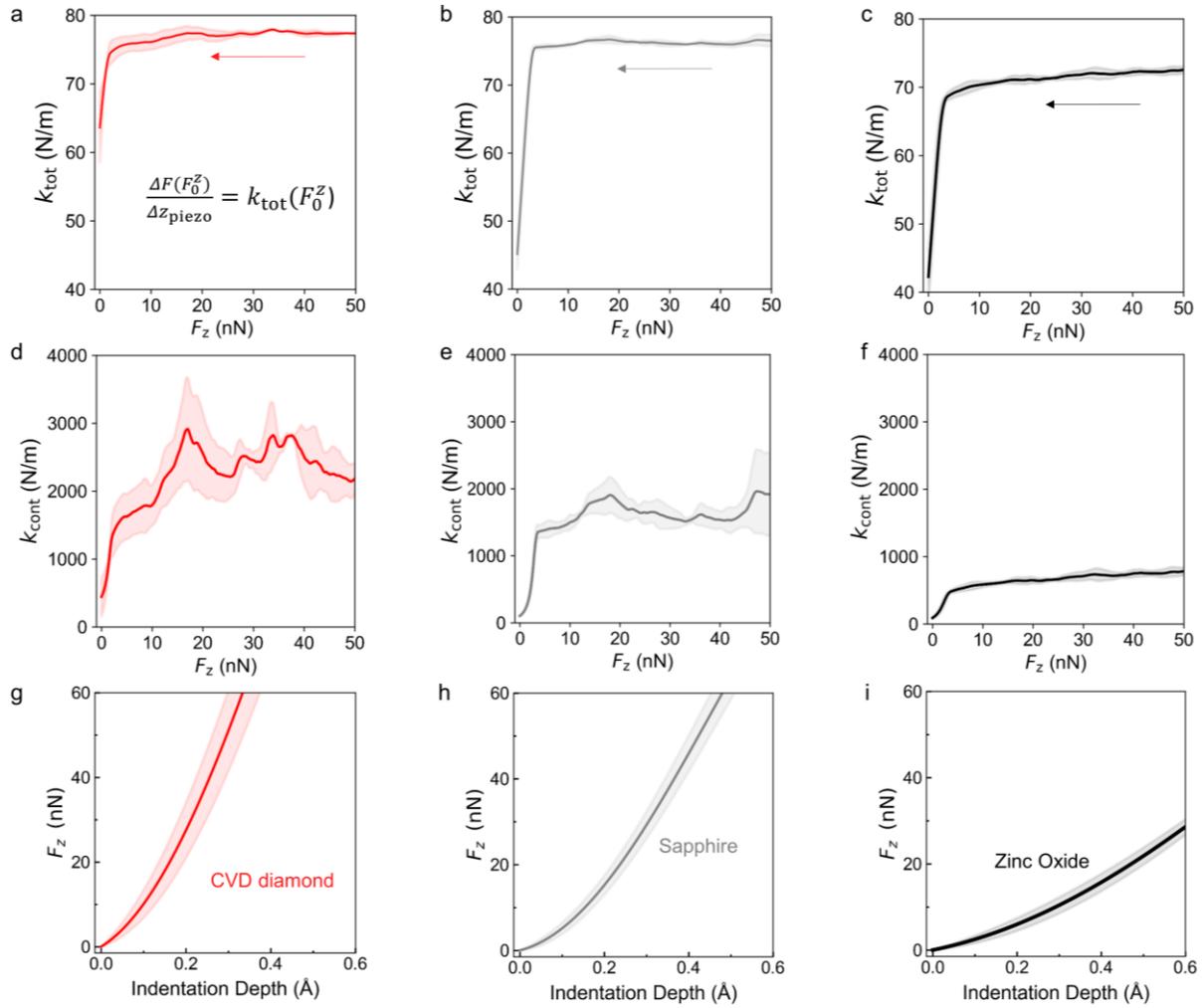

**Figure 8. Experimental MoNI/ÅI indentation measurements for different substrates.** Experimental total stiffness and contact stiffness (solid lines) for (a,d) CVD diamond, (b,e) sapphire, and (c,f) ZnO. (d-f) The associated indentation curves computed using equation (5) for (g) CVD diamond, (h) Polycrystalline sapphire, and (i) ZnO. Solid lines are mean curves computed over at least 5 different positions for each samples. Shaded area is one standard deviation from the mean.

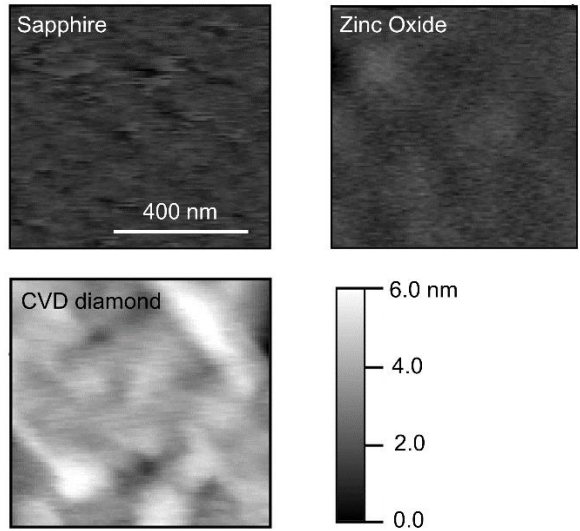

**Figure 9. MoNI/ÅI samples topography.** Topographies of the surface of the sapphire, ZnO, and CVD diamond samples employed in the experiments.

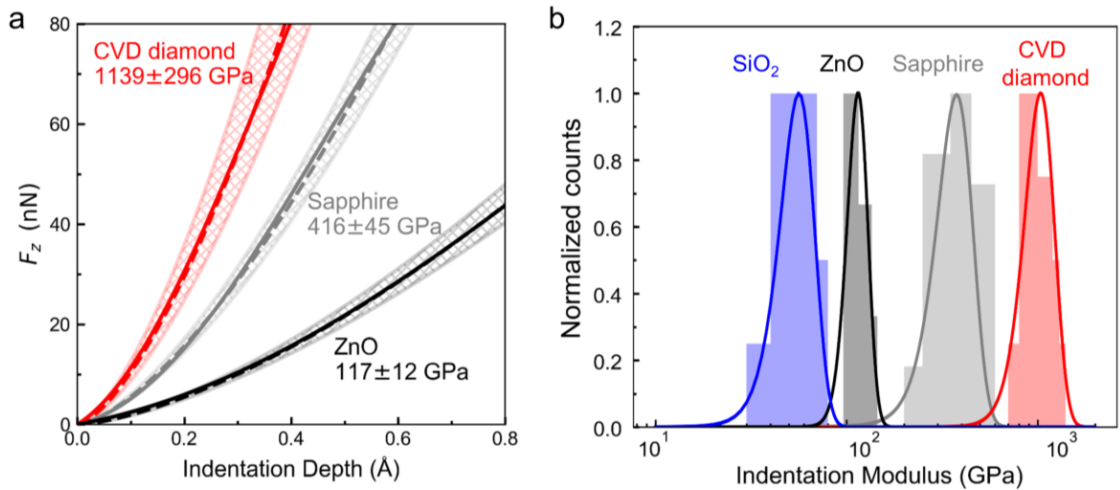

**Figure 10. MoNI/ÅI indentation curves data fitting.** (a) Experimental indentation curves computed using equation (4) for ZnO (black solid line), sapphire (gray solid line), and CVD diamond (red solid line). The dashed lines are the result of the nonlinear fitting with equation (5). The mean value of the modulus identified through the fitting procedure is reported in the figure for the three materials, together with the associated standard deviation. The shaded areas correspond to one standard deviation from the mean of the fitted curves. (b) Normalized distributions of indentation moduli for ZnO (115 ± 14 GPa), Sapphire (387 ± 81 GPa), CVD diamond (1005 ± 188 GPa) and $SiO_2$ (56 ± 11 GPa, the relatively high variance on the value of the indentation modulus of $SiO_2$ is likely associated to the presence of adsorbates on the surface). Data are obtained by aggregate data for 15-30 experiments for each material.
35

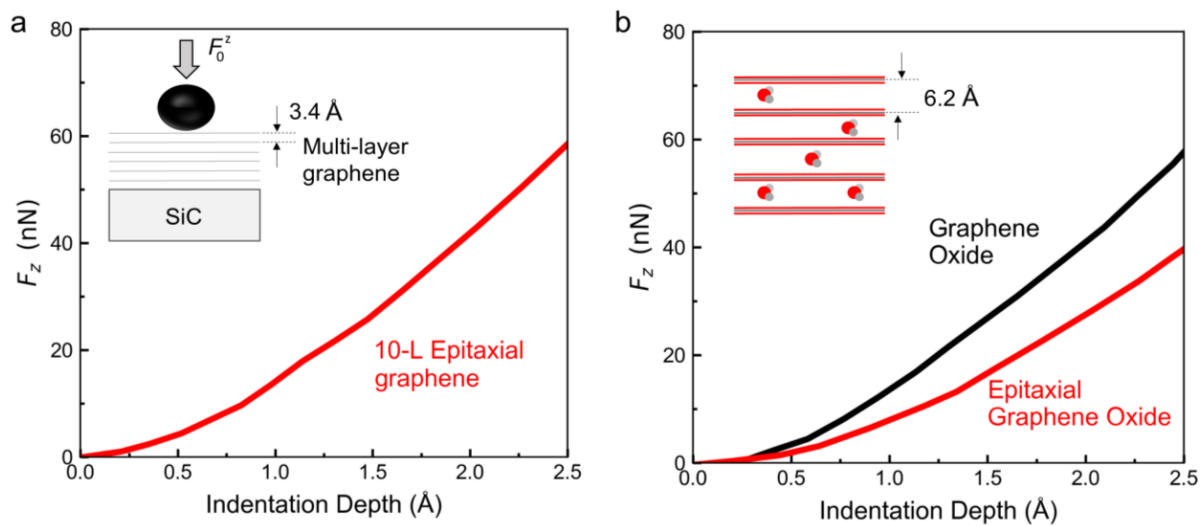

**Figure 11. MoNI/ÅI indentation of graphene.** (a) Experimental indentation curve for 10-L epitaxial graphene on Silicon Carbide (SiC). In the graphic, schematic of the indentation problem for the graphene sample. (b) Experimental indentation curve for graphene oxide on Silicon and epitaxial graphene oxide on Silicon Carbide. In the graphic, the schematic of layers distribution and the intercalation of water molecules in graphene oxide interlayers.

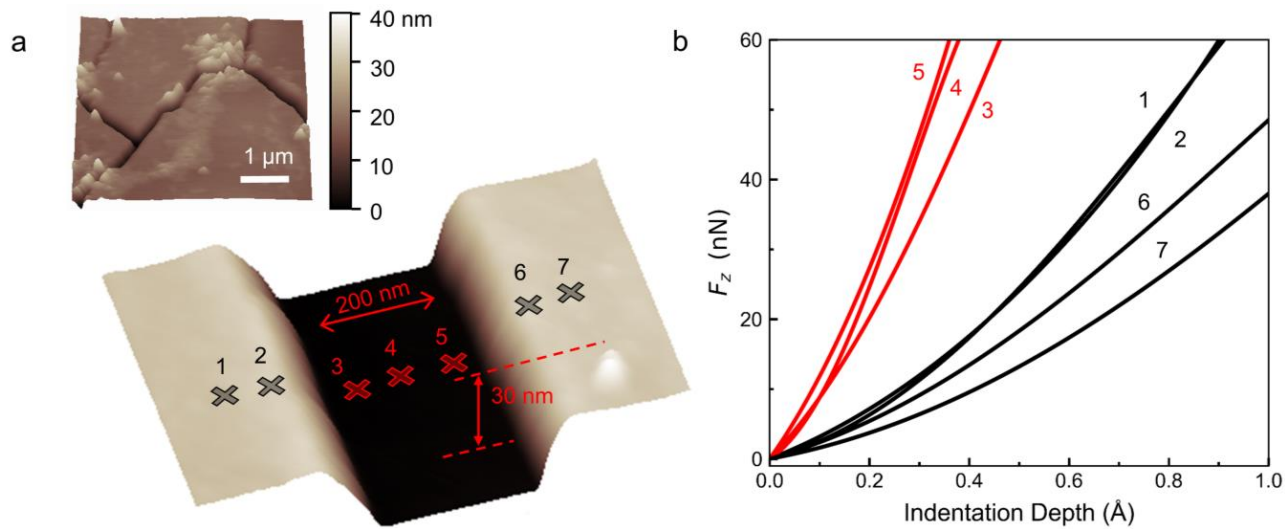

**Figure 12. MoNI/ÅI indentation of Q-carbon nanostructures.** (a) Bottom image: topography of the section of a Q-carbon filament (see also Reference [37]), the crosses indicate the position where MoNI/ÅI is performed. Top image, topography of the surface of a Q-carbon sample. (b) Indentation curves for the Q-carbon filament (red lines) and the surrounding diamond-like-carbon material. The numbers refer to the positions indicated in (a).